\documentclass[twocolumn,showpacs,aps,prd]{revtex4-1}
\usepackage{amsmath}
\usepackage{dcolumn}
\usepackage{graphicx}
\usepackage{subfigure}
\usepackage{epstopdf}
\usepackage{color}
\usepackage{multirow}
\usepackage{setspace}
\usepackage{overpic}
\usepackage{amssymb}
\usepackage[colorlinks,urlcolor=blue, citecolor=blue,linkcolor=blue] {hyperref}
\usepackage{lineno}
\usepackage{bm}
\usepackage{rotating}
\usepackage[utf8]{inputenc}
\let\oldequation\equation
\let\oldendequation\endequation
\renewenvironment{equation}
 {\linenomathNonumbers\oldequation}
 {\oldendequation\endlinenomath}

\newcommand{\Jeeppp}{J/\psi \to e^+ e^- \pi^0 \pi^+ \pi^-}

\begin{document}

\title{\bf \boldmath
Observation of the decay $J/\psi \to e^+ e^- \eta(1405)$ with $\eta(1405) \to \pi^0 f_0(980)$
}
\date{\today}

\author{
\begin{small}
\begin{center}
M.~Ablikim$^{1}$, M.~N.~Achasov$^{5,b}$, P.~Adlarson$^{75}$, X.~C.~Ai$^{81}$, R.~Aliberti$^{36}$, A.~Amoroso$^{74A,74C}$, M.~R.~An$^{40}$, Q.~An$^{71,58}$, Y.~Bai$^{57}$, O.~Bakina$^{37}$
% ...
\\
\vspace{0.2cm}
(BESIII Collaboration)\\
\vspace{0.2cm} {\it
$^{1}$ Institute of High Energy Physics, Beijing 100049, People's Republic of China\\
% ...
$^{5}$ Budker Institute of Nuclear Physics SB RAS (BINP), Novosibirsk 630090, Russia\\
% ...
$^{36}$ Johannes Gutenberg University of Mainz, Johann-Joachim-Becher-Weg 45, D-55099 Mainz, Germany\\
$^{37}$ Joint Institute for Nuclear Research, 141980 Dubna, Moscow region, Russia\\
% ...
$^{40}$ Liaoning Normal University, Dalian 116029, People's Republic of China\\
% ...
$^{57}$ Southeast University, Nanjing 211100, People's Republic of China\\
$^{58}$ State Key Laboratory of Particle Detection and Electronics, Beijing 100049, Hefei 230026, People's Republic of China\\
% ...
$^{71}$ University of Science and Technology of China, Hefei 230026, People's Republic of China\\
% ...
$^{74}$ University of Turin and INFN, (A)University of Turin, I-10125, Turin, Italy; (B)University of Eastern Piedmont, I-15121, Alessandria, Italy; (C)INFN, I-10125, Turin, Italy\\
$^{75}$ Uppsala University, Box 516, SE-75120 Uppsala, Sweden\\
% ...
$^{81}$ Zhengzhou University, Zhengzhou 450001, People's Republic of China\\

\vspace{0.2cm}
% ...
$^{b}$ Also at the Novosibirsk State University, Novosibirsk, 630090, Russia\\
% ...

}\end{center}

\vspace{0.4cm}
\end{small}
}

\begin{abstract}
Using a data sample of $(10087\pm44)\times 10^6$ $J/\psi$ events collected by the BESIII detector in 2009, 2012, 2018 and 2019, the electromagnetic Dalitz process $J/\psi \to e^+ e^- \eta(1405)$
is observed via the decay $\eta(1405) \to \pi^0 f_0(980)$, $f_0(980) \to \pi^+ \pi^-$, with a significance of about $9.6\sigma$.
The branching fraction of this decay is measured to be
${\mathcal B}(J/\psi \to e^+ e^- \pi^0 \eta(1405) \to e^+ e^- \pi^0 f_0(980) \to e^+ e^- \pi^0 \pi^+ \pi^-)=(2.02\pm0.24(\rm{stat.})\pm0.09(\rm{syst.}))\times 10^{-7}$.
The branching-fraction ratio ${\mathcal B}(J/\psi \to e^+ e^- \eta(1405))$/${\mathcal B}(J/\psi \to \gamma \eta(1405))$
is determined to be $(1.35\pm0.19(\rm{stat.})\pm0.06(\rm{syst.}))\times10^{-2}$.
Furthermore, an $e^+e^-$ invariant-mass dependent transition form factor of $J/\psi \to e^+ e^-\eta(1405)$ is presented for the first time. The obtained result provides input for different theoretical models, and is valuable for the improved understanding the intrinsic structure of the $\eta(1405)$ meson.
\end{abstract}

\maketitle

\oddsidemargin  -0.2cm
\evensidemargin -0.2cm

\section{\boldmath INTRODUCTION}
The $\eta(1405)$ and $\eta(1475)$ excited isoscalar states with $J^{PC}=0^{-+}$ were discovered in the $K\bar{K}\pi$ invariant mass distribution in $p\bar{p}$ collisions in 1967~\cite{1}. Since then, several theoretical and experimental studies have been performed to investigate the nature of these states. They are considered to be possible candidates for pseudoscalar glueballs~\cite{2}, however their measured masses are much lower than those calculated by lattice Quantum Chromodynamics~(QCD), which are greater than 2~GeV$/c^2$~\cite{3_1,3_2,3_3,3_4}. Further studies from several experiments (i.e. E769~\cite{14051475_1}, MARKIII~\cite{14051475_2}, L3~\cite{14051475_3}, E852 (MPS)~\cite{14051475_4}, DM2~\cite{14051475_5,14051475_6}, OBELIX~\cite{14051475_7, 14051475_8, 14051475_9, 14051475_10}) indicated that
the two pseudoscalar states, $\eta(1405)$ and $\eta(1475)$, have rather different properties. The $\eta(1405)$ state has large couplings with $a_0(980)\pi$ or directly with $K\bar{K}\pi$; it can also decay into $\eta \pi^+ \pi^-$, while the $\eta(1475)$ state decays mainly to $K^{*}\bar{K}$.
Furthermore, there remain unsolved puzzles in the understanding of their production mechanism and decay properties,
and it can not yet be concluded whether $\eta(1405)$ and $\eta(1475)$ are two separate states or a single state observed in different decay modes~\cite{4}.
In 2012, the isospin violating process $\eta(1405) \to f_0(980) \pi^0$ was observed in the decay $J/\psi \to \gamma \eta(1405) \to \gamma \pi^0 f_0(980) \to \gamma \pi^0 \pi^+ \pi^-$~\cite{f0width}.
Following this measurement, several theoretical papers~\cite{JJWu2012, XGWu2013, mengchun} demonstrated that the ``triangle singularity'' mechanism~(TSM) could lead to the anomalously large isospin violation effect, and also deform the line shapes and shift the peak positions of the $\eta(1405/1475)$. Hence, further studies of $\eta(1405/1475) \to \pi^0 \pi^+ \pi^-$ are imperative and will be valuable for better understanding the $\eta(1405/1475)$ puzzle and study the properties of these states.

The study of electromagnetic~(EM) Dalitz decays plays an important role in the understanding of the intrinsic structure of hadrons and the interaction between photons and hadrons.
In such a process, a virtual photon converts into a lepton pair, and the four-momentum transferred by the virtual photon corresponds to the invariant mass of the lepton pair.
The branching fraction (BF) ratios of the EM Dalitz decays to the corresponding radiative decays of vector mesons are suppressed by two orders of magnitude~\cite{Rpee,ppiee,wpiee,Retaee}.

The EM properties of a $J/\psi$ to pseudoscalar transition are characterized by the EM transition form factor~(TFF), which can be calculated in QCD models~\cite{TFFQCD1,TFFQCD2,TFFQCD3}. In experiments, the TFF can be directly determined by comparing the measured invariant-mass spectrum of the lepton pair from the EM Dalitz decay with the Quantum Electrodynamics~(QED) prediction for point-like particles.
The $q^2$-dependent differential BF for $J/\psi\to e^+ e^- \eta(1405)$, normalized to the corresponding radiative decay $J/\psi\to \gamma \eta(1405)$, is given by
\begin{small}
\begin{equation}
\frac{d {\mathcal B}\left(J / \psi \rightarrow e^{+} e^{-}\eta(1405)\right)}{d q^2 {\mathcal B}(J / \psi \rightarrow \gamma\eta(1405) )}=\left|F\left(q^2\right)\right|^2 \times\left[\operatorname{QED}\left(q^2\right)\right],
\end{equation}
\end{small}where the squared four-momentum transfer $q^2$ is equal to the square of the invariant mass of the $e^+e^-$ pair, $\left|F\left(q^2\right)\right|^2$ is the normalized TFF for the $J/\psi\to \eta(1405)$ transition, and $\operatorname{QED}\left(q^2\right)$ represents the point-like QED result~\cite{qedTFF}.
Measurements of the TFF may lead to a better understanding of charmonium Dalitz decays and their internal structure,  and provide tests of QCD predictions.

This paper reports the first observation of the EM Dalitz decay channel of $\Jeeppp$ via $\eta(1405)$ in the $\pi^0 \pi^+ \pi^-$ invariant mass spectrum and the measurement of the TFF of $J/\psi \to e^+ e^- \eta(1405)$.
The ratio, $R=\frac{{\mathcal B}(J/\psi \to e^+ e^- \eta(1405))}{{\mathcal B}(J/\psi \to \gamma \eta(1405))}$,
is evaluated with inputs from a previous study of the radiative decay~\cite{f0width}.

\section{\boldmath{BESIII Detector and Data Set}}\label{detectormcsample}
The BESIII detector~\cite{Ablikim:2009aa} records symmetric $e^+e^-$ collisions
provided by the BEPCII storage ring~\cite{Yu:IPAC2016-TUYA01}
in the center-of-mass energy range from 2.0 to 4.95~GeV,
with a peak luminosity of $1 \times 10^{33}\;\text{cm}^{-2}\text{s}^{-1}$
achieved at $\sqrt{s} = 3.77\;\text{GeV}$.
BESIII has collected large data samples in this energy region~\cite{Ablikim:2019hff}. The cylindrical core of the BESIII detector covers 93\% of the full solid angle and consists of a helium-based
 multilayer drift chamber~(MDC), a plastic scintillator time-of-flight
system~(TOF), and a CsI(Tl) electromagnetic calorimeter~(EMC),
which are all enclosed in a superconducting solenoidal magnet
providing a 1.0~T (0.9~T in
2012, for about $(1088.5\pm4.4)\times10^6$ $J/\psi$ events~\cite{jpsinumber}) magnetic field. The solenoid is supported by an
octagonal flux-return yoke with resistive plate counter muon
identification modules interleaved with steel.
The charged-particle momentum resolution at $1~{\rm GeV}/c$ is
$0.5\%$, and the specific energy deposits (d$E$/d$x$) resolution is $6\%$ for electrons
from Bhabha scattering. The EMC measures photon energies with a
resolution of $2.5\%$ ($5\%$) at $1$~GeV in the barrel (end-cap)
region. The time resolution in the TOF barrel region is 68~ps, while
that in the end-cap region is 110~ps. The end-cap TOF system was upgraded
in 2015 with multi-gap resistive plate chamber technology, providing a
time resolution of 60 ps~\cite{MRPC1, MRPC2}, which is used for a data
set of $(8774.0\pm39.4)\times10^6$ $J/\psi$ events~\cite{jpsinumber} taken
in 2017-2019.

This analysis uses a data sample of $(10087\pm44)\times 10^6$ $J/\psi$ events~\cite{jpsinumber} accumulated in $e^+ e^-$ collisions at a center-of-mass energy $\sqrt{s} = 3.097$~GeV by BESIII. Monte-Carlo~(MC) simulation of the full detector is used to determine the detection
efficiency, optimize event selection criteria, and estimate
backgrounds. The simulation program {\sc boost}~\cite{boost} provides an event generator,
contains the detector geometry description~\cite{detvis}, and simulates the detector response
and signal digitization. The production of the $J/\psi$ resonance is simulated by
the MC event generator {\sc kkmc}~\cite{kkmc1,kkmc2}, while the known decay modes are generated with {\sc evtgen}~\cite{besevtgen1,besevtgen2} using branching fractions from the
PDG~\cite{brpsip}, and the remaining unknown charmonium decays are modelled with {\sc lundcharm}~\cite{ref:lundcharm_1, ref:lundcharm_2}.

For the signal MC sample, the simulation of the $J/\psi \to e^+ e^- \eta(1405)$ channel takes into account the angular distributions of the final states and the polarization of the $J/\psi$~\cite{JPLL}. The subsequent decay $\eta(1405) \to \pi^0 \pi^+ \pi^-$ is modelled by a custom generator based on a theoretical model~\cite{mengchun} with modifications to incorporate the TSM effects~\cite{TS11, TS12} due to the intermediate $K^*\bar{K} + c.c.$ rescatterings by exchanging an on-shell kaon (anti-kaon) in the decay of $\eta(1405) \to \pi^0 \pi^+ \pi^-$. The width effect within the triangle loops and the isospin violation arising
from the $K^0/K^{\pm}$ mass difference within triangle loops are included.
The masses of $K^0$, $K^{\pm}$ and the mass and width of $K^{*0}$ are set to the world-average values~\cite{brpsip} in the simulation.

\section{\boldmath{EVENT SELECTION}}
Charged tracks detected in the MDC are required to be within a polar angle ($\theta$) range of $|\!\cos\theta|<0.93$, where $\theta$ is defined with respect to the $z$-axis, which is the symmetry axis of the MDC.
The point of closest approach to the interaction point (IP) must have the distance to the IP less than 10\,cm along the $z$-axis
and less than 1\,cm in the transverse plane. The number of good charged tracks is required to be equal to four with zero net charge. Particle identification~(PID) for charged tracks combines measurements of the specific ionization energy loss in the MDC~(d$E$/d$x$), the flight time in the TOF and the energy deposited in the EMC to form likelihoods for electron
[$\mathcal{L}(e)$], pion [$\mathcal{L}(\pi)$], and kaon [$\mathcal{L}(K)$] hypotheses. Tracks are identified as electrons when the electron hypothesis has the highest PID likelihood among the three hypotheses, i.e. $\mathcal{L}(e)>\mathcal{L}(K)$ and $\mathcal{L}(e)>\mathcal{L}(\pi)$.
To exclude misidentified pions, an additional $E/p > 0.8$ requirement is applied to the $e^+/e^-$ tracks with momentum higher than 0.8 GeV/$c$, where $E$ and $p$ are the energy deposited in the EMC and the momentum measured with the MDC, respectively.
Any charged track not identified as an electron is assigned as a pion. An event is accepted if it contains exactly the track combination $e^+e^-\pi^+\pi^-$.

Photon candidates are identified using showers in the EMC. The deposited energy of each shower must be more than 25 MeV in the barrel EMC ($|\!\cos\theta| < 0.8$) or 50 MeV in the end-cap EMC $(0.86 < |\!\cos\theta| < 0.92)$. To exclude showers that originate from charged particles, the angle subtended by the EMC shower and the extrapolated position of the closest charged track at the EMC must be greater than 10 degrees. To suppress electronic noise and showers unrelated to the event, the difference between the EMC time and the event start time is required to be within
[0, 700]\,ns. At least two photons are retained in each event.

A vertex fit is performed to the four charged tracks~($e^+e^- \pi^+ \pi^-$) to ensure that they originate from the same point. An energy-momentum constraint (4C) kinematic fit is also performed for the selected $~J/\psi \to e^+e^-\eta(1405) \to e^+ e^- \pi^0 f_0(980) \to e^+ e^- \pi^0 \pi^+ \pi^-$ candidate events.
For events with more than two photon candidates, the combination with the least $\chi^2_{\mathrm{4C}}$ from the kinematic fit is retained.
The requirement of $\chi^2_{\mathrm{4C}}$ is optimized based on the figure-of-merit $S/\sqrt{S+B}$, where $S$ and $B$ stand for the number of signal and background events passing the above selection criteria, respectively. Finally, the $\chi^2_{\mathrm{4C}}$ value is required to be less than 20.

A Crystal Ball function is used to describe the distribution of invariant masses of the selected photon pairs ($M_{\gamma\gamma}$).
The pair is accepted as a $\pi^0$ candidate if the invariant mass $M_{\gamma\gamma}$ is within a
$\pm 3\sigma$ mass window around the $\pi^0$ peak, where $\sigma$ is the resolution of the fitted Crystal Ball function.

\begin{figure}[htbp]
\centering
  \includegraphics[width=7cm, height=4.5cm]{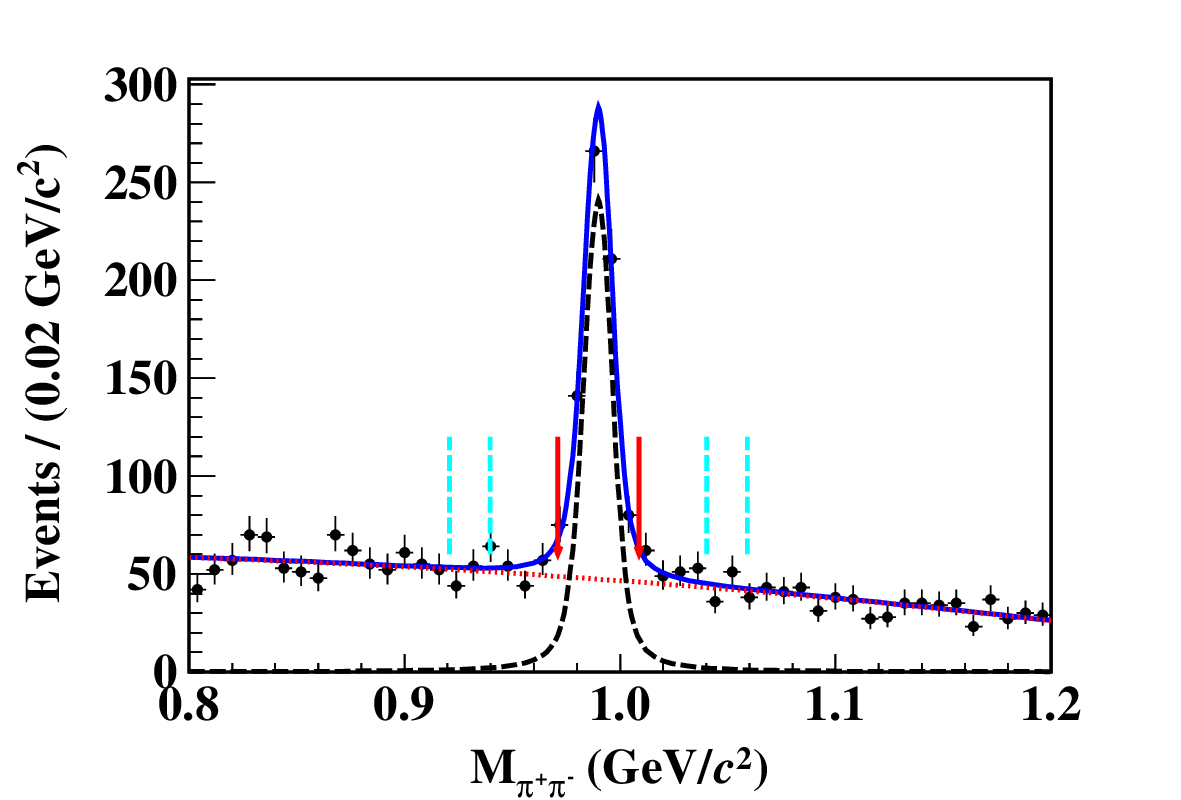}
  \caption{Distribution of $M_{\pi^+\pi^-}$ for the accepted candidates with fit results overlaid. The dots with error bars represent data, the blue solid curve is the overall fit, the black dashed line is the fitted signal, the red dotted line represents the fitted background. The red arrows indicate the $f_0(980)$
mass window~($0.971~\rm{GeV}$/$c^2 < M_{\pi^+\pi^-} < 1.009~\rm{GeV}$/$c^2$), and the pairs of cyan dashed vertical lines (left and right of the signal peak) indicate the $f_0(980)$ sideband regions~($0.921~\rm{GeV}$/$c^2 < M_{\pi^+\pi^-} < 0.940~\rm{GeV}$/$c^2$ and $1.040~\rm{GeV}$/$c^2 < M_{\pi^+\pi^-} < 1.059~\rm{GeV}$/$c^2$).}
\label{Jpsitoeeeta_pipi_fit}
\end{figure}

The events are required to have the $\pi^+\pi^-$ invariant mass~($M_{\pi^+\pi^-}$) consistent with the world average
mass of the $f_0(980)$ meson\cite{brpsip}. To define the $M_{\pi^+\pi^-}$ selection window, a fit to the $M_{\pi^+\pi^-}$
distribution is performed in the signal-enriched range of the $\pi^+ \pi^- \pi^0$ invariant mass
between 1.3 and 1.5~GeV/$c^2$~\cite{f0width}. The signal is described by a convolution
of a Breit-Wigner function and a Gaussian resolution function while the background shape
is described by a second-order Chebychev polynomial.
The mass and width of the Breit-Wigner function
and the Gaussian resolution are free parameters of the fit.
The events with $M_{\pi^+\pi^-}$ within $\pm 3\sigma$ around the $f_0(980)$ peak are kept for the further analysis,
where $\sigma$ is the width of the Breit-Wigner function. The fit result is shown in Fig.~\ref{Jpsitoeeeta_pipi_fit}.

Events in which a photon converts into an $e^+e^-$ pair in the
material of the detector constitute a source of background, in
particular those from $J/\psi \to \gamma \eta(1405)$ decays.
The distance $R_{xy}$ from the IP to the reconstructed point of closest approach of the $e^+e^-$ pair is used to separate the signal events from the photon-conversion background events~\cite{Rxyconv}. For the signal process the $e^+ e^-$ pair comes directly from the $J/\psi$ decay, so $R_{xy}$ values are close to zero. Background events accumulate above 2~cm, where most of the photon conversions occur. To suppress this background the value of $R_{xy}$ is required to be less than 2~cm.

Possible background contributions are studied with a sample of 2932 pb$^{-1}$ of data taken at $\sqrt s =$~3.773~GeV~\cite{data_3773} and the $J/\psi$ inclusive MC sample. The analysis of the 3.773~GeV sample indicates that the QED background is negligible, while studies of the MC sample reveal no significant sources of peaking background.

\section{\boldmath Branching-fraction measurement}
An unbinned maximum-likelihood fit is performed on the $M_{\pi^+\pi^-\pi^0}$ spectrum  in the $f_0(980)$ signal
and sideband regions simultaneously. After taking into account the mass dependence of the detection efficiency,
a convolution of a Breit-Wigner function with a Gaussian function is used as the $\eta(1405)$ signal shape.
The mass and width are free parameters in the fit, while the Gaussian resolution is taken from the MC. The small bump around 1.3~GeV$/c^2$ is described by a Breit-Wigner function,
the mass and width of which are fixed at the world-average values of the $f_1(1285)$ resonance~\cite{brpsip}. The background is described by a third-order
Chebychev function with parameters determined by a simultaneous fit to the $M_{\pi^+\pi^-\pi^0}$ spectrum with $M_{\pi^+\pi^-}$ being
in the $f_0(980)$ sidebands region~($0.921~\rm{GeV}$/$c^2 < M_{\pi^+\pi^-} < 0.940~\rm{GeV}$/$c^2$ and $1.040~\rm{GeV}$/$c^2 < M_{\pi^+\pi^-} < 1.059~\rm{GeV}$/$c^2$). The yield of each component is a free parameter of the fit.

\begin{figure}[htbp]
\centering
   \includegraphics[width=7.5cm, height=6cm]{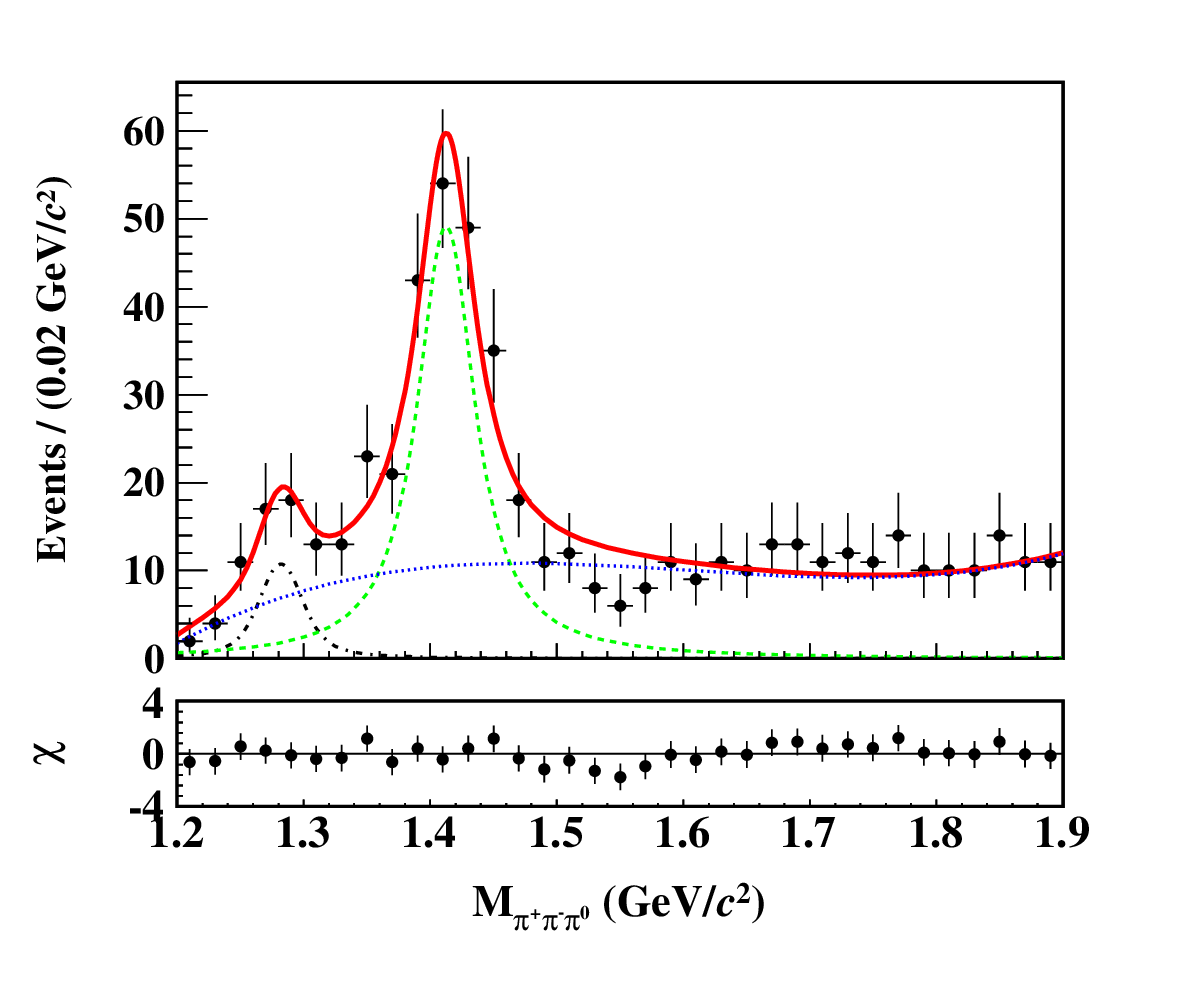}
  \caption{Top: Distribution of $M_{\pi^+\pi^-\pi^0}$ for the accepted candidates with $M_{\pi^+\pi^-}$ being in the $f_0(980)$ signal region with fit results overlaid. The black dots with error bars represent data, the red solid curve is the overall fit, the black dash-dotted line denotes the $f_1(1285)$ signal, the green dashed line denotes the $\eta(1405)$ signal and the blue dotted line is background. Bottom: The pull between the fit and data,
\emph{i.e.} the difference between data and the fit curve
normalised by the uncertainty of the data in each bin.}
\label{Jpsitoeeeta fitdata}
\end{figure}

The fit result is shown in Fig.~\ref{Jpsitoeeeta fitdata}.  The number of $\eta(1405)$ signal events is $194.2\pm23.1$. The detection efficiency is estimated to be $(9.64\pm0.05)\%$ using the signal MC simulation.
The significance of the $\eta(1405)$ signal peak is $9.6\sigma$, calculated from the difference of the likelihood values with and without the $\eta(1405)$ signal in the fits. The calculation of the significance takes into account the systematic effects discussed later.
The measured BF of the decay $J/\psi \to e^+ e^- \eta(1405) \to e^+ e^- \pi^0 f_0(980) \to e^+ e^- \pi^0 \pi^+ \pi^-$ is calculated as
\begin{equation}
\begin{split}
\label{Jpsitoeeeta formula}
  {\mathcal B}_{\rm sig} = \frac{N_{\rm sig}}{N_{J/\psi} \cdot \epsilon \cdot \mathcal{B}(\pi^0 \to \gamma \gamma)},
\end{split}
\end{equation}
where $N_{\rm sig}$ is the number of signal events, $N_{J/\psi}$ is the total number of
$J/\psi$ events~\cite{jpsinumber}, $\epsilon$ is the detection
efficiency and $\mathcal{B}(\pi^0 \to \gamma \gamma)$
is the BF of $\pi^0\to \gamma\gamma$ \cite{brpsip}.
We obtain ${\mathcal B}_\text{sig} = (2.02 \pm 0.24) \times 10^{-7}$
where the uncertainty is statistical only.

\section{\boldmath branching-fraction systematic uncertainties}
The sources of systematic uncertainties for the BF measurement include the MDC tracking efficiency, the PID efficiency, the $N_{\rm track}$ requirement, the photon detection efficiency, the $\gamma$ conversion veto, the total number of $J/\psi$ events, the BF of $\pi^0 \to \gamma \gamma$, the kinematic fit, the $M_{\pi^+\pi^-}$ window, the $\pi^0$ mass window, the fit range of $M_{\pi^+\pi^-\pi^0}$ and the background shape.

To study the systematic uncertainties related to the MDC tracking and PID efficiencies of $e^\pm$, the differences between data and MC simulation in different momentum and polar-angle
regions are studied  with a control sample of radiative Bhabha scattering $e^+ e^- \to \gamma e^+ e^-$~(including $J/\psi \to \gamma e^+ e^-$) events from the same data sample.
The associated systematic uncertainties are determined by applying bin-by-bin weights, which take into account the data/MC differences
and the uncertainty due to the limited size of the control sample,
to the momentum and polar-angle distributions of $e^\pm$ in the signal decays.
The systematic uncertainties from $e^\pm$ tracking and PID are 0.4\% and 0.5\%, respectively.

The systematic uncertainty due to the $E/p>0.8$ requirement is estimated with
the same control sample and a weighting method similar to that used in the PID efficiency study. The difference in the efficiencies of the $E/p>0.8$ requirement is tabulated in bins of high momentum and polar angle.
The systematic uncertainty for the $E/p > 0.8$ requirement is determined to be 0.1\%.

The systematic uncertainties from the tracking efficiency of charged pions and the photon-detection efficiency are investigated by analyzing  control samples of $J/\psi \to \pi^+ \pi^- \pi^0$ and $e^+e^- \to \gamma \mu^+ \mu^-$ events, respectively.
The two-dimensional distributions of the polar angle and momentum in the signal decays are reweighted according
to the observed data/MC differences and the statistical uncertainties of the control samples.
The systematic uncertainty of the $\pi^\pm$ tracking efficiency is assigned to be 0.6\% per track and that of photon detection efficiency is 0.5\% per photon.

The control sample of $J/\psi \to \pi^+ \pi^- \pi^0$, $\pi^0 \to \gamma e^+ e^-$ events is used to study the systematic bias associated with  the requirement of $N_{\rm track} = 4$. The difference in the ratio of $N_{\rm track} = 4$ to $N_{\rm track} \geq 4$ between data and MC simulation is considered as a systematic uncertainty, which is 0.2\%.

The systematic uncertainty from the photon conversion veto criterion $R_{xy} < 2$~cm is assessed with the same control sample of $J/\psi \to \pi^+\pi^-\pi^0$, $\pi^0 \to \gamma e^+e^-$ events. The relative difference of efficiencies associated with the $\gamma$ conversion rejection between data and MC simulation, is 0.12\%~\cite{nGood:sys}, which is taken as the associated systematic uncertainty.
The MC samples of the radiation decay $J/\psi \to \gamma \eta(1405)$ are generated to estimate the systematic
 uncertainty from the remaining gamma conversion background in the data. The difference of events between the data and MC, 0.4\%, is taken as the systematic uncertainty.

The track helix parameters are corrected for simulated events to reduce the difference between data and MC simulation in the $\chi^2_{\mathrm{4C}}$ distribution. Half of the difference in the selection efficiencies after and before the correction, 0.7\%, is taken as the corresponding
systematic uncertainty~\cite{kinematic_fit}.

In the baseline fit, the resonances $f_1(1285)$ and $\eta(1405)$ are described by an incoherent sum of Breit-Wigner functions,
designated as $BW_1$ and $BW_2$, respectively. To estimate the systematic uncertainty from neglected interference effects, the probability density function~(PDF) of the two resonances is constructed as
\begin{linenomath*}
\begin{align}
\rm{PDF}_{\text {signal}} &=(|A \times B W_{1}(s)     \nonumber\\
&+ BW_{2}(s) \times e^{i \varphi}|^{2}) \otimes G(s) \times \varepsilon(s),
\end{align}\end{linenomath*}
where $\varphi$ is the relative phase between the $\eta(1405)$ and $f_1(1285)$, $A$ is the normalization factor, $G(s)$ is a Gaussian function used to describe the data-MC difference on the detector resolution, and $\varepsilon(s)$ is the detection efficiency curve obtained by the MC simulation. The difference between the fits with and without interference, 2.1\%, is taken to be the systematic uncertainty.

The signal region in the $M_{\pi^+\pi^-}$ distribution is widened by the width of the Breit-Wigner function that describes the distributions, and another fit is performed.   The difference on the measured BF is taken as the corresponding systematic uncertainty, which is 1.5\%.

The fit range of $M_{\pi^+ \pi^- \pi^0}$ is changed from the baseline values [1.2, 1.9]~GeV/$c^2$ to [1.2, 1.84]~GeV/$c^2$ and [1.2, 1.96]~GeV/$c^2$
and the  $f_0(980)$ sideband range is changed to ($0.915~\rm{GeV}$/$c^2 < M_{\pi^+\pi^-} < 0.940~\rm{GeV}$/$c^2$ and $1.040~\rm{GeV}$/$c^2 < M_{\pi^+\pi^-} < 1.065~\rm{GeV}$/$c^2$).
The largest variation on the measured BF between the alternative fits and the baseline fit, 2.0\%, is assigned as the associated systematic uncertainty.

The signal MC sample is used to derive an alternative shape for the signal distribution, which is then included in a new fit the $M_{\gamma\gamma}$ distribution. The relative difference on the number of signal events, 1.0\%, is taken as the systematic uncertainty.

The uncertainty from the TFF is estimated with the alternative signal
MC samples generated with the pole mass $\Lambda =$ 1.96, 3.097 or 3.77~GeV$/c^2$ instead of the default value 3.686~GeV$/c^2$, and
the largest charge of the efficiency, 1.7\%, is taken as the systematic uncertainty.
To determine the systematic uncertainty due to the custom generator, the known width of $K^{*}$ is changed by $\pm 1\sigma$ or $\pm 2\sigma$ to produce different signal MC samples.
The largest change of the efficiency, which is 0.4\%, is added in quadrature to the pole-mass uncertainty to arrive at a total uncertainty of  1.8\%, which is considered to be the systematic uncertainty associated with the custom generator.

The uncertainty due to the BF of $\pi^0 \to \gamma \gamma$ is 0.034\%~\cite{brpsip}.
The systematic uncertainty on the total number of $J/\psi$ events is 0.44\% according to Ref.~\cite{jpsinumber}.

All the sources of systematic uncertainties are summarized in Table~\ref{systematic}. The total systematic uncertainty is evaluated to be 4.3\% from the quadratic sum of the above individual contributions.

\begin{table}[htbp]
\begin{center}
 \caption{Relative systematic uncertainties in the branching-fraction measurement.}
  \label{systematic}
  \renewcommand{\arraystretch}{1.0}
  \vspace{0.2cm}
  \begin{tabular}{cc}
  \hline\hline
  Source          &Uncertainty~(\%)   \\\hline
  $e^\pm$ tracking &0.4 \\
  $e^\pm$ PID      &0.5   \\
  $E/p>0.8$ requirement   & 0.1   \\
  $\pi^\pm$ tracking  &1.2  \\
  Photon detection   &1.0  \\
  $N_{\rm track} = 4$     &0.2    \\
  Veto of $\gamma$ conversion &0.2 \\
  $\gamma$ conversion background &0.4 \\
  4C kinematic fit        &0.7  \\
  Interference effects among resonances   &2.1 \\
  $f_0(980)$ mass window   &1.5  \\
  Fit range and background shape &2.0 \\
 $\pi^0$ mass window     &1.0  \\
  Custom generator  &1.8     \\
  ${\mathcal B}(\pi^0 \to \gamma \gamma)$   &$<$0.1\,\,\,  \\
  $N_{J/\psi}$    &0.5   \\
  Total      &4.3 \\
  \hline\hline
  \end{tabular}
  \vspace{-0.2cm}
  \end{center}
\end{table}

\section{\boldmath Transition Form Factor Measurement}
\begin{figure}[htbp]
\centering
   \includegraphics[width=0.4\textwidth]{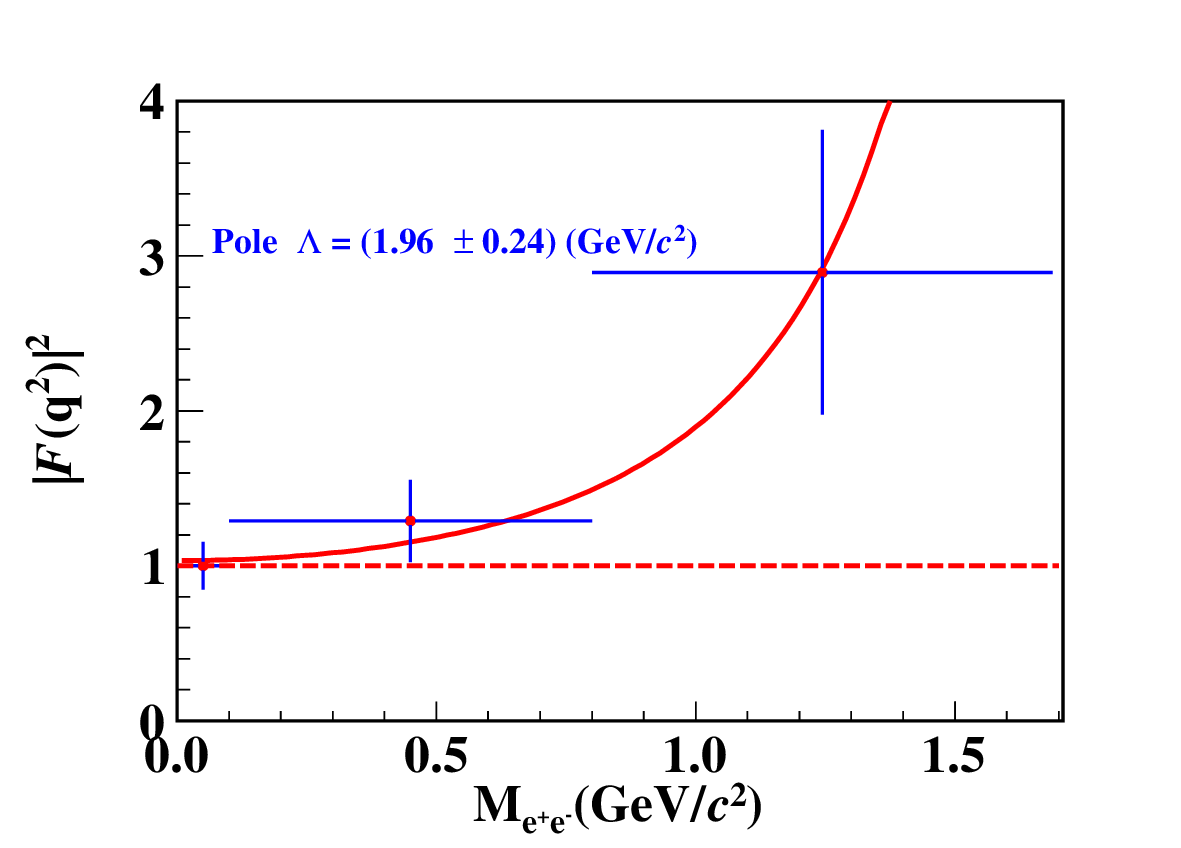}
  \caption{The TFF distribution for the $J/\psi \to e^+ e^- \eta(1405)$ decay.
The dots with error bars denote the $|\mathcal{F}(q^2)|^2$ values, the red solid curve is the result of the fit by the simple pole approximation and the red dash line shows $|\mathcal{F}(q^2)|^2 = 1$.}
\label{Jpsitoeeeta TFF}
\end{figure}

The TFF for $J/\psi \to e^+ e^- \eta(1405)$ decay is measured by dividing the $M_{e^+e^-}$ spectrum into three intervals according to the ratio of the measured BFs and the BFs predicted by QED. The former are obtained using the same selection criteria and fit method as for the BF measurement, and the latter is calculated by Eq.~12 of Ref.~\cite{qedTFF} in each bin.
The normalized TFF is parameterized in the usual simple pole approximation as~\cite{qedTFF}
\begin{eqnarray}
\label{Jpsitoeeeta Fq}
\mathcal{F}(q^{2}) = \frac{1}{1-q^2/\Lambda ^2},
\end{eqnarray}
where $q^2=M^{2}_{e^+e^-}$ is the square of the invariant mass of $e^+e^-$ pair, and the parameter $\Lambda$ is the spectroscopic pole
mass, which is used as a fit parameter. Figure~\ref{Jpsitoeeeta TFF}
shows the result of the fit to the $M_{e^+e^-}$ dependence of $|\!\mathcal{F}(q^2)\!|^2$. From the fit, $\Lambda$ is determined to be $(1.96 \pm 0.24 (\rm{stat.}) \pm 0.06(\rm{syst.}))~\rm{GeV}$/$c^2$.
The systematic uncertainty on the $\Lambda$ measurement is obtained by repeating the fit after considering variations similar to those used in the determination of the systematic uncertainty of the BF measurement.

\section{\boldmath SUMMARY}\label{summary}
Based on the data sample of $(10087\pm44)\times 10^6$ $J/\psi$ events collected with the BESIII in 2009, 2012, 2018 and 2019, the EM Dalitz decay of $J/\psi \to e^+ e^- \eta(1405)$ is observed for the first time via $\eta(1405) \to \pi^0 f_0(980)$ and $f_0(980) \to \pi^+ \pi^-$ with a significance of $9.6\sigma$. The BF of $J/\psi \to e^+ e^- \eta(1405) \to e^+ e^- \pi^0 f_0(980) \to e^+ e^- \pi^0 \pi^+ \pi^-$ is measured to be $(2.02\pm0.24(\rm{stat.})\pm0.09(\rm{syst.}))\times10^{-7}$, where the first uncertainty is statistical and the second one is systematic. Considering the measured BF of the radiative decay of $J/\psi \to \gamma \eta(1405) \to \gamma \pi^0 f_0(980) \to \gamma \pi^0 \pi^+ \pi^-$~\cite{f0width},
the BF ratio $\frac{{\mathcal B}(J/\psi \to e^+ e^- \eta(1405))}{{\mathcal B}(J/\psi \to \gamma \eta(1405))}$ is calculated to be $(1.35\pm0.19(\rm{stat.})\pm0.06(\rm{syst.}))\times10^{-2}$, against which future theoretical calculations can be compared.
The TFF for the $J/\psi \to e^+ e^- \eta(1405)$ decay is extracted by assuming the simple pole approximation. The pole value in the vector meson dominance model, $\Lambda =(1.96 \pm 0.24 (\rm{stat.}) \pm 0.06(\rm{syst.}))$~GeV/$c^2$ is also measured. The obtained result provides valuable input for different theoretical models, and understanding the intrinsic structure of $\eta(1405)$ mesons.

\vspace{8pt}
\section*{\boldmath ACKNOWLEDGEMENTS}
The BESIII Collaboration thanks the staff of BEPCII and the IHEP computing center for their strong support. This work is supported in part by National Key R\&D Program of China under Contracts Nos. 2020YFA0406300, 2020YFA0406400; National Natural Science Foundation of China (NSFC) under Contracts Nos. 11635010, 11735014, 11835012, 11935015, 11935016, 11935018, 11961141012, 12022510, 12025502, 12035009, 12035013, 12061131003, 12192260, 12192261, 12192262, 12192263, 12192264, 12192265, 12221005, 12225509, 12235017; the Chinese Academy of Sciences (CAS) Large-Scale Scientific Facility Program; the CAS Center for Excellence in Particle Physics (CCEPP); CAS Key Research Program of Frontier Sciences under Contracts Nos. QYZDJ-SSW-SLH003, QYZDJ-SSW-SLH040; 100 Talents Program of CAS; The Institute of Nuclear and Particle Physics (INPAC) and Shanghai Key Laboratory for Particle Physics and Cosmology; ERC under Contract No. 758462; European Union's Horizon 2020 research and innovation programme under Marie Sklodowska-Curie grant agreement under Contract No. 894790; German Research Foundation DFG under Contracts Nos. 443159800, 455635585, Collaborative Research Center CRC 1044, FOR5327, GRK 2149; Istituto Nazionale di Fisica Nucleare, Italy; Ministry of Development of Turkey under Contract No. DPT2006K-120470; National Research Foundation of Korea under Contract No. NRF-2022R1A2C1092335; National Science and Technology fund of Mongolia; National Science Research and Innovation Fund (NSRF) via the Program Management Unit for Human Resources \& Institutional Development, Research and Innovation of Thailand under Contract No. B16F640076; Polish National Science Centre under Contract No. 2019/35/O/ST2/02907; The Swedish Research Council; U. S. Department of Energy under Contract No. DE-FG02-05ER41374.



\begin{thebibliography}{**}

\bibitem{1} P. H. Baillon {\it et al.},
  \href{https://link.springer.com/content/pdf/10.1007/BF02823526.pdf}{Nuovo Cimento {\bf 50A}, 393 (1967)}.

  \bibitem{2} L. Faddeev, A. J. Niemi, and U. Wiedner,
  \href{https://doi.org/10.1103/PhysRevD.70.114033}{Phys. Rev. D {\bf 70}, 114033 (2004).}

  \bibitem{3_1} G. S. Bali {\it et al.} (UKQCD Collaboration),
  \href{https://doi.org/10.1016/0370-2693(93)90948-H}{Phys. Lett. B {\bf 309}, 378 (1993)}.

  \bibitem{3_2} C. J. Morningstar and M. J. Peardon,
  \href{https://doi.org/10.1103/PhysRevD.60.034509}{Phys. Rev. D {\bf 60}, 034509 (1999)}.

  \bibitem{3_3} Y. Chen et al.,
  \href{https://doi.org/10.1103/PhysRevD.73.014516}{Phys. Rev. D {\bf 73}, 014516 (2006)}.

  \bibitem{3_4}  A. Chowdhury, A. Harindranath, and J. Maiti,
  \href{https://doi.org/10.1103/PhysRevD.91.074507}{Phys. Rev. D {\bf 91}, 074507 (2015)}.

   \bibitem{14051475_1} M. G. Rath {\it et al.},
  \href{https://doi.org/10.1103/PhysRevD.40.693}{Phys. Rev. D {\bf 40}, 693 (1989)}.

  \bibitem{14051475_2} Z. Bai {\it et al.} (Mark III Collaboration),
  \href{https://doi.org/10.1103/PhysRevLett.65.2507}{Phys. Rev. Lett. {\bf 65}, 2507 (1990)}.

  \bibitem{14051475_3} M. Acciarri {\it et al.} (L3 Collaboration),
  \href{https://doi.org/10.1016/S0370-2693(01)00102-2}{Phys. Lett. B {\bf 501}, 1 (2001)}.

  \bibitem{14051475_4} Adams {\it et al.},
  \href{https://doi.org/10.1016/S0370-2693(01)00951-0}{Phys. Lett. B {\bf 516}, 264 (2001)}.

  \bibitem{14051475_5} J. E. Augustin {\it et al.} (DM2 Collaboration),
  \href{https://doi.org/10.1103/PhysRevD.42.10}{Phys. Rev. D {\bf 42}, 10 (1990)}.

  \bibitem{14051475_6} J. E. Augustin {\it et al.} (DM2 Collaboration),
  \href{https://doi.org/10.1103/PhysRevD.46.1951}{Phys. Rev. D {\bf 46}, 1951 (1992)}.



  \bibitem{14051475_7} A. Bertin {\it et al.} (Obelix Collaboration),
  \href{https://doi.org/10.1016/0370-2693(95)01136-E}{Phys. Lett. B {\bf 361}, 187 (1995)}.

  \bibitem{14051475_8} C. Cical\` o {\it et al.} (Obelix Collaboration),
  \href{https://doi.org/10.1016/S0370-2693(99)00898-9}{Phys. Lett. B {\bf 462}, 453 (1999)}.

  \bibitem{14051475_9} A. Bertin {\it et al.} (Obelix Collaboration),
  \href{https://doi.org/10.1016/S0370-2693(97)00300-6}{Phys. Lett. B {\bf 400}, 226 (1997)}.

  \bibitem{14051475_10} F. Nichitiu {\it et al.} (Obelix Collaboration),
  \href{https://doi.org/10.1016/S0370-2693(02)02547-9}{Phys. Lett. B {\bf 545}, 261 (2002)}.



  \bibitem{4} E. Klempt and A. Zaitsev,
  \href{https://doi.org/10.1016/j.physrep.2007.07.006}{Phys. Rep. \bf {454}, 1 (2007)}.

  \bibitem{f0width} M. Ablikim {\it et al.} (BESIII Collaboration),
  \href{https://doi.org/10.1103/PhysRevLett.108.182001}{Phys. Rev. Lett. {\bf 108}, 182001 (2012)}.

  \bibitem{JJWu2012} J. J. Wu, X. H. Liu, Q. Zhao, and B. S. Zou,
  \href{https://doi.org/10.1103/PhysRevLett.108.081803}{Phys. Rev. Lett. {\bf 108}, 081803 (2012)}.

  \bibitem{XGWu2013} X. G. Wu, J. J. Wu, Q. Zhao, and B. S. Zou,
  \href{https://doi.org/10.1103/PhysRevD.87.014023}{Phys. Rev. D {\bf 87}, 014023 (2013)}.

  \bibitem{mengchun} M. C. Du and Q. Zhao,
  \href{https://doi.org/10.1103/PhysRevD.100.036005}{Phys. Rev. D {\bf 100}, 036005~(2019)}.

  \bibitem{Rpee} M. Ablikim {\it et al.} (BESIII Collaboration),
  \href{https://doi.org/10.1103/PhysRevD.89.092008}{Phys. Rev. D {\bf 89}, 092008 (2014)}.

  \bibitem{ppiee} A. Anastasi {\it et al.} (KLOE-2 Collaboration),
  \href{https://doi.org/10.1016/j.physletb.2016.04.015}{Phys. Lett. B {\bf 757}, 362~(2016)}.

  \bibitem{wpiee} P. Adlarson {\it et al.} (A2 Collaboration),
  \href{https://doi.org/10.1103/PhysRevC.95.035208}{Phys. Rev. C {\bf 95}, 035208~(2017)}.

  \bibitem{Retaee} M. Ablikim {\it et al.} (BESIII Collaboration),
  \href{https://doi.org/10.1016/j.physletb.2018.05.038}{Phys. Lett. B {\bf 783}, 452 (2018)}.

  \bibitem{TFFQCD1} F. Klingl, N. Kaiser, and W. Weise,
  \href{https://doi.org/10.1007/BF02769217}{Z. Phys. A {\bf 356}, 193 (1996)}.

  \bibitem{TFFQCD2} A. Faessler, C. Fuchs, and M. I. Krivoruchenko,
  \href{https://doi.org/10.1103/PhysRevC.61.035206}{Phys. Rev. C {\bf 61}, 035206 (2000)}.

  \bibitem{TFFQCD3} C. Terschlusen and S. Leupold,
  \href{https://doi.org/10.1016/j.physletb.2010.06.033}{Phys. Lett. B {\bf 691}, 191 (2010)}.

  \bibitem{qedTFF} L. M. Gu, H. B. Li, X. X. Ma, and M. Z. Yang,
  \href{https://doi.org/10.1103/PhysRevD.100.016018}{Phys. Rev. D {\bf 100}, 016018 (2019)}.

  \bibitem{Ablikim:2009aa}
  M.~Ablikim {\it et al.} (BESIII Collaboration),
  \href{https://doi.org/10.1016/j.nima.2009.12.050}{Nucl.\ Instrum.\ Meth.\ A {\bf 614}, 345 (2010)}.

 \bibitem{Yu:IPAC2016-TUYA01} C.~H.~Yu {\it et al.},
  \href{https://doi.org/10.18429/JACoW-IPAC2016-TUYA01}{Proceedings of IPAC2016, Busan, Korea, 2016}.

  \bibitem{Ablikim:2019hff}
  M.~Ablikim {\it et al.} (BESIII Collaboration),
  \href{https://doi.org/10.1088/1674-1137/44/4/040001}{Chin. Phys. C {\bf 44}, 040001 (2020)}.

\bibitem{jpsinumber} M. Ablikim {\it et al.} (BESIII Collaboration),
      \href{https://doi.org/10.1088/1674-1137/ac5c2e}{Chin. Phys. C {\bf 46}, 074001 (2022)}.

  \bibitem{MRPC1} X. Li {\it et al.},
  \href{https://link.springer.com/article/10.1007\%2Fs41605-017-0012-4}{Rad. Det. Tech. Meth. {\bf 1}, 13~(2017)}.

  \bibitem{MRPC2} Y. X. Guo {\it et al.},
  \href{https://link.springer.com/article/10.1007\%2Fs41605-017-0012-4}{Rad. Det. Tech. Meth. {\bf 1}, 15~(2017)}.


  \bibitem{boost} Z. Y. Deng {\it et al.},
  \href{http://hepnp.ihep.ac.cn/article/id/283d17c0-e8fa-4ad7-bfe3-92095466def1}{Chin. Phys. C {\bf 30}, 371 (2006).}

   \bibitem{detvis} K.~X.~Huang, {\it et al.},
   \href{https://doi.org/10.1007/s41365-022-01133-8}{Nucl.\ Sci.\ Tech. {\bf 33}, 142 (2022).}


  \bibitem{kkmc1} S. Jadach, B. F. L. Ward, and Z. Was,
  \href{https://doi.org/10.1016/S0010-4655(00)00048-5}{Comput. Phys. Commun. {\bf 130}, 260 (2000).}

  \bibitem{kkmc2} S. Jadach, B. F. L. Ward, and Z. Was,
  \href{https://doi.org/10.1103/PhysRevD.63.113009}{Phys. Rev. D {\bf 63}, 113009 (2001).}

  \bibitem{besevtgen1} D. J. Lange,
  \href{https://doi.org/10.1016/S0168-9002(01)00089-4}{Nucl. Instrum. Meth., Sect. A {\bf 462}, 152 (2001).}

  \bibitem{besevtgen2} R.~G. Ping,
  \href{https://doi.org/10.1088/1674-1137/32/8/001}{Chin. Phys. C {\bf 32}, 599~(2008).}

  \bibitem{brpsip} R.L. Workman \textit{et al.} (Particle Data Group),
  \href{https://dx.doi.org/10.1093/ptep/ptac097}{Prog. Theor. Exp. Phys. {\bf 2022}, 083C01 (2022)}.

  \bibitem{ref:lundcharm_1} J.~C.~Chen, G.~S.~Huang, X.~R.~Qi, D.~H.~Zhang and Y.~S.~Zhu,
  \href{https://doi.org/10.1103/PhysRevD.62.034003}{Phys.\ Rev.\ D {\bf 62}, 034003 (2000).}
  \bibitem{ref:lundcharm_2} R.~L.~Yang, R.~G.~Ping and H.~Chen,
  \href{https://doi.org/10.1088/0256-307X/31/6/061301}{Chin.\ Phys.\ Lett.\  {\bf 31}, 061301 (2014).}

  \bibitem{JPLL} J. L. Fu, H. B. Li, X. S. Qin, and M. Z. Yang,
  \href{https://doi.org/10.1142/S0217732312502239}{Mod. Phys. Lett. A {\bf 27}, 1250223~(2012).}

  \bibitem{TS11} L. D. Landau,
  \href{https://doi.org/10.1016/0029-5582(59)90154-3}{Nucl. Phys., {\bf 13}, 1 (1959)}.

  \bibitem{TS12} R. E. Cutkosky,
  \href{https://doi.org/10.1063/1.1703676}{J. Math. Phys., {\bf 1}, 429 (1960)}.

  \bibitem{Rxyconv} Z. R. Xu and K. L. He,
   \href{https://doi.org/10.1088/1674-1137/36/8/010}{Chin. Phys. C {\bf 36}, 742~(2012)}.

   \bibitem{data_3773} M. Ablikim \textit{et al.} (BESIII Collaboration),
  \href{https://doi.org/10.1088/1674-1137/37/12/123001}{Chin. Phys. C {\bf 37}, 123001~(2013)};
  \href{https://doi.org/10.1016/j.physletb.2015.11.043}{Phys. Lett. B 753, 629 (2016)}.


  \bibitem{nGood:sys} M. Ablikim {\it et al.} (BESIII Collaboration),
  \href{https://doi.org/10.1103/PhysRevLett.129.022002}{Phys. Rev. Lett. {\bf 129}, 022002 (2022)}.

  \bibitem{kinematic_fit} M. Ablikim {\it et al.} (BESIII Collaboration),
  \href{https://journals.aps.org/prd/pdf/10.1103/PhysRevD.87.012002}{Phys. Rev. D {\bf 87}, 012002 (2013)}.


\end{thebibliography}
\end{document}